\documentclass[twocolumn,showpacs,preprintnumbers,amsmath,amssymb,floatfix]{revtex4}
\usepackage{graphicx}
\usepackage{dcolumn}
\usepackage{bm}
\usepackage{graphicx}
\usepackage{color}
\usepackage{upgreek}

\begin{document}
\preprint{APS/123-QED}

\title{A spallation-based neutron target for direct studies of neutron-induced reactions in inverse kinematics}

\author{Ren{\'e}~Reifarth}
\email{reifarth@physik.uni-frankfurt.de}
\affiliation{Goethe-Universit\"{a}t Frankfurt, Frankfurt am Main, Germany}
\author{Kathrin~G\"obel}
\affiliation{Goethe-Universit\"{a}t Frankfurt, Frankfurt am Main, Germany}
\author{Tanja~Heftrich}
\affiliation{Goethe-Universit\"{a}t Frankfurt, Frankfurt am Main, Germany}
\author{Beatriz~Jurado}
\affiliation{CENBG, Gradignan, France}
\author{Franz~K\"appeler}
\affiliation{Karlsruhe Institute of Technology, Karlsruhe, Germany}
\author{Yuri A. Litvinov}
\affiliation{GSI Helmholtzzentrum f\"ur Schwerionenforschung, Darmstadt, Germany}
\author{Mario Weigand}
\affiliation{Goethe-Universit\"{a}t Frankfurt, Frankfurt am Main, Germany}

\date{\today}

\begin{abstract}

We discuss the possibility to build a neutron target for nuclear reaction studies in inverse kinematics utilizing a storage ring and radioactive ion beams. The proposed neutron target is a specially designed spallation target surrounded by a large moderator of heavy water (D$_2$O). We present the resulting neutron spectra and their properties as a target. We discuss possible realizations at different experimental facilities. 
\end{abstract}

\pacs{25.40.Lw, 26.20.+f, 28.41.-i, 29.20.db }


\maketitle

\section{Introduction}
\label{intro}

Neutron capture cross sections of unstable isotopes are important for neutron induced nucleosynthesis as well as for technological applications. The traditional time-of-flight method~\cite{RLK14} reaches its limits once the necessary detection of the reaction products is hampered by the size (mass) of the sample. Several factors may limit the sample mass: (a) The decay properties of radioactive isotopes interfere with the signals from the neutron capture or neutron-induced fission reactions~\cite{CoR07}. (b) The limited range of charged reaction products requires a thin sample. In both cases, an increased neutron fluence at the sample position with ever improved neutron sources overcomes the lack of reaction rate~\cite{HRF01,RHH04,RBA04}.

\begin{figure}[!h]
\begin{center}
\renewcommand{\baselinestretch}{1}
\small\normalsize
  \includegraphics[width=.49\textwidth]{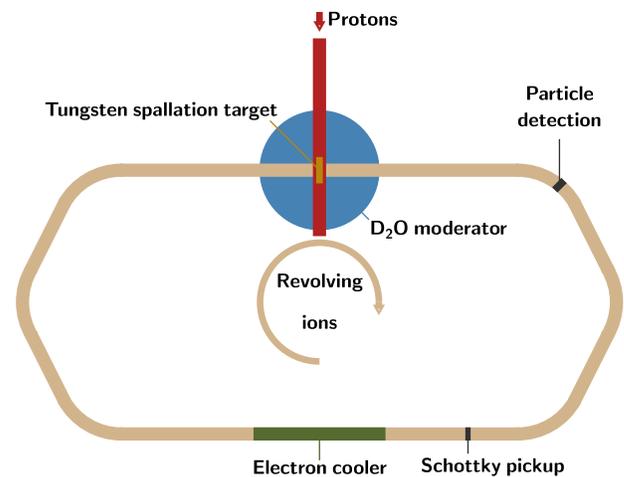}
  \caption{\label{fig:rough_setup} Rough sketch of the proposed setup. Neutrons are produced by protons impinging on a tungsten spallation target (brown). The proton beam pipe (red) is orientated perpendicular to the ion beam pipe (light brown). The beam pipes do not intersect. The neutrons produced in the spallation process are moderated in the surrounding heavy-water (blue). They penetrate the ion beam pipe and act as a neutron target for the ions. The ion beam pipe is part of a storage ring outside the moderator. The storage ring may contain additional equipment like an electron cooler (green), Schottky pickups and particle detectors (gray).}
\end{center}
\end{figure}

Reference~\cite{ReL14} proposed a combination of a radioactive beam facility, an ion storage ring and a high flux reactor to allow direct measurements of neutron-induced reactions over a wide energy range on isotopes with half lives down to minutes. The authors discussed specific reactions, detection techniques and counting rates. Here, we present the possibility to replace the rather demanding reactor with a specially designed spallation neutron source. FIG.~\ref{fig:rough_setup} shows a sketch of the proposed setup. 

The advantages of such a setup over the reactor approach are manifold: No critical assembly is required, and therefore, the safety and security regulations are much less stringent. No actinides at all are used or produced. In particular, no minor actinides are produced avoiding long-lived radioactive waste. Last but not least, there are considerably less $\gamma$-rays per neutron.

Similar neutron densities as in a research reactor can be reached in a close-by ion beam pipe if the spallation target is surrounded by a moderator of heavy water (D$_2$O). We present the concept in Section~\ref{scheme} and the corresponding simulations in Section~\ref{simulations}. It is feasible to build such a neutron target at facilities like LANSCE at LANL (USA), n\_TOF/ISOLDE at CERN (Switzerland), GSI/FAIR (Germany), HIRFL-CSR/HIAF (China), and others. We discuss the possible realizations in Section~\ref{realizations}.

\section{Concept and geometry}
\label{scheme}

\begin{figure}
\begin{center}
\renewcommand{\baselinestretch}{1}
\small\normalsize
  \includegraphics[width=.9\textwidth]{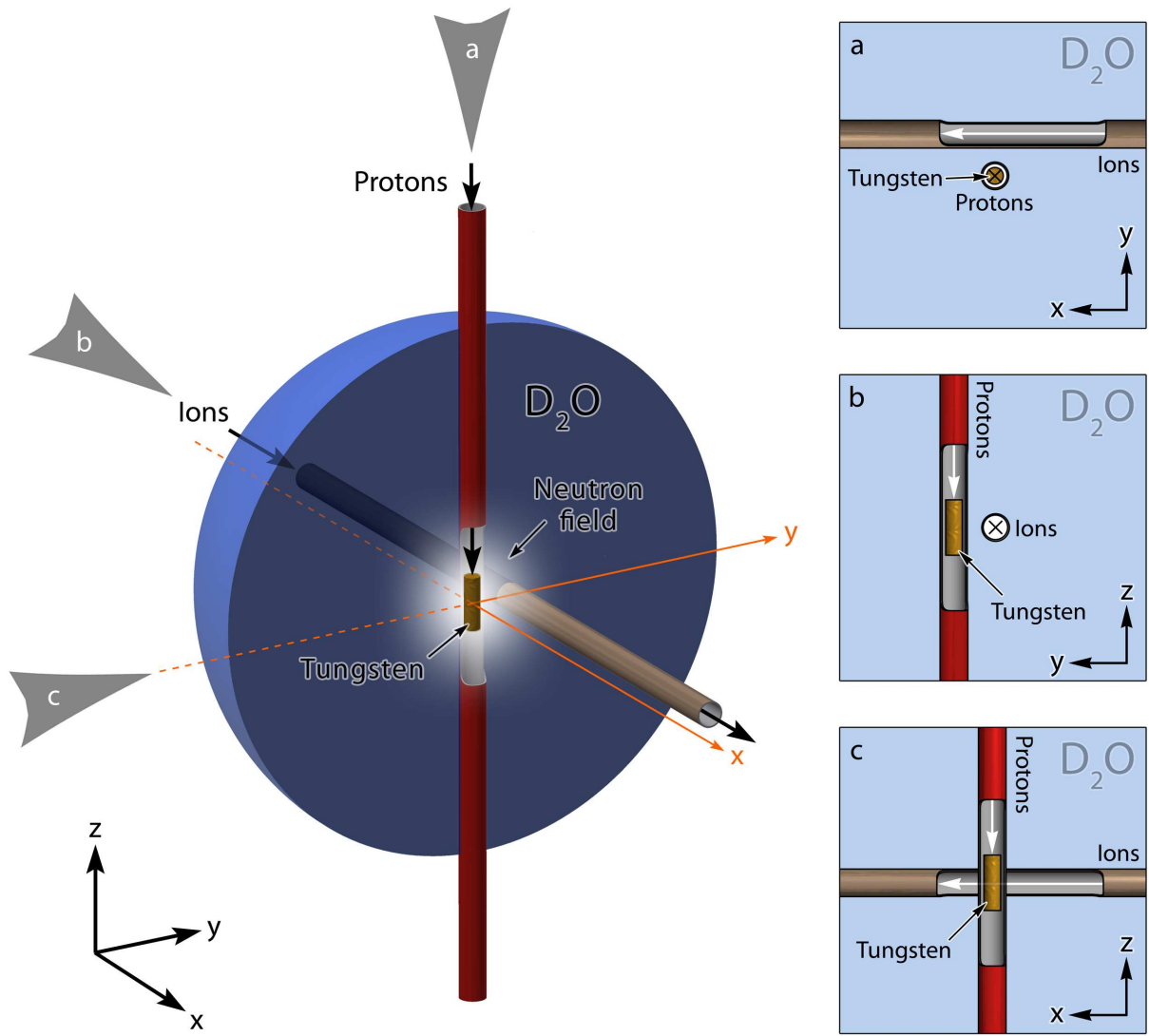}
  \caption{\label{fig:setup} Sketch of the proposed setup. Left: Protons impinge on a tungsten cylinder and produce neutrons. The tungsten spallation target (brown) inside the propton beam pipe (red) is shown in the center of the sketch. (In the sketch, the pipe is cut open to reveal the tungsten target.) The proton beam pipe points along the z-axis and is orientated perpendicular to the ion beam pipe (gray). The beam pipes do not intersect as the ion beam pipe is shifted by several centimeters. The neutrons produced in the spallation process are moderated in the surrounding heavy-water (blue, only half of the sphere is shown). They penetrate the ion beam pipe and act as a neutron target for the ions. The ends of ion beam pipe can be connected on the outside to form a storage ring. Right: 2D-projections of the setup (a) along the proton beam pipe, (b) along the ion beam pipe, and (c) perpendicular to both pipes. The lines of sight are indicated by gray arrows in the left sketch. The water moderator is indicated by the light blue background.}
\end{center}
\end{figure}

The center of the simulated setup is a tungsten spallation target, see FIG.~\ref{fig:setup}. The cylindrical target is mounted inside an evacuated proton beam pipe with a radius of 2.5~cm and aligned in the direction of the proton beam. The protons impinge on the tungsten and produce neutrons. The material of the beam pipe has to be chosen such that it has only minor effects on the neutrons. The neutrons are moderated outside the proton beam pipe by heavy water in a surrounding sphere. A second (ion) beam pipe is orientated perpendicular to the proton beam pipe. The two pipes do not intersect since the ion beam pipe is shifted by x~=~7.5 cm off the center of the setup. The neutrons penetrate into the ion beam pipe and serve as a target for the ions. 

Compared to other elements, tungsten provides a high density of about 19~g~cm$^{-3}$ combined with a very high melting point of about 3,700~K. We investigated two different tungsten target sizes, the ''small version'' with a radius of 1.5~cm and a length of 10~cm, and the ''large version'' with a radius of 2.5~cm and length of 50~cm. 

We chose heavy-water moderators of different sizes with radii 0.5~m, 1.0~m, and 2.0~m. If not specified otherwise, the center of the spherical moderator is also the center of the spallation target. A technical realization would require a cooling of the spallation target. The cooling should be realized with heavy water to avoid changes in the neutron physics investigated here. Most neutrons eventually escape the moderator volume after moderation and have to be absorbed on the outside. Again, this will not alter the neutron budget.

\clearpage

\onecolumngrid

\twocolumngrid

\section{Simulations}\label{simulations}

We simulated the proposed setup with GEANT-3.21 \cite{GEA93} with the GCALOR package \cite{ZeG94}. As a first step, neutrons of different energies were emitted isotropically from the center of the tungsten target with the tungsten target in the center of the moderator (section~\ref{neutrons}). These simulations help to understand the underlying principles and allow first rough estimates of the neutron density in the ion beam pipe. In addition, they can be used to estimate the effect of different primary neutron energies. In the second step, we simulated the interaction of a high-energy proton beam with a tungsten target (section~\ref{protons}). All particles were followed until they either left the volume or were absorbed.

\subsection{Neutrons started with different energies \label{neutrons}}

Neutrons with different energies from $10^{-2}$ to $10^{9}$~eV were randomly started in the tungsten spallation target for each moderator thickness (radii 0.5~m, 1.0~m, and 2.0~m). The direction of the neutrons was chosen isotropically. A separate simulation for each energy decade was performed. The energy distribution within a decade was assumed to be $1/E$~\cite{RBA04}. Each simulation run included $10^{6}$ neutrons. 

The neutrons penetrate the ion beam pipe where they act as a neutron target for the ions. We investigated the average time period a neutron spends in the ion beam pipe. The average time period depends on the velocity of the neutron, the passing angle, and the number of times the neutron actually crosses the ion beam pipe. The time period spent inside the ion tube is summed up for each neutron emitted from the spallation target and stored into a histogram.

\begin{figure}
\begin{center}
\renewcommand{\baselinestretch}{1}
\small\normalsize
  \includegraphics[width=.49\textwidth]{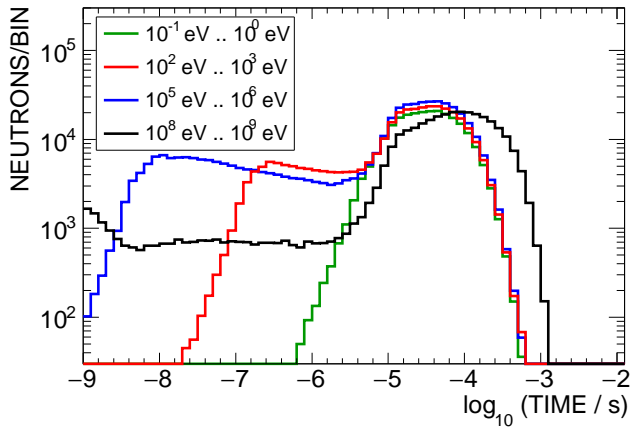}
  \caption{\label{fig:log_times} Total time periods neutrons spend inside the ion beam pipe which is surrounded by a moderator of 1~m radius. The different histograms correspond to different primary neutron energy ranges, which were chosen exemplary from the entire neutron energy range from $10^{-2}$ to $10^{9}$~eV. For each energy decade, $10^6$~initial neutrons were simulated. Events where the neutron bypasses the ion beam pipe are not shown (zero-suppressed). About 30\% to 50\% of all neutrons pass the ion beam pipe at least once. The x-axis has a logarithmic binning with 10~bins per time decade.}
\end{center}
\end{figure}

FIG.~\ref{fig:log_times} shows examples for neutrons of selected primary energy ranges. When the primary neutron energy increases, the number of events with at least one pass through the beam pipe increases as well as the number of crossings. In addition, very short time period of a few 10$^{-9}$~s are only observed for higher primary neutron energies. At lower energies even a single pass requires a few 10$^{-6}$~s.


\begin{figure}
\begin{center}
\renewcommand{\baselinestretch}{1}
\small\normalsize
  \includegraphics[width=.49\textwidth]{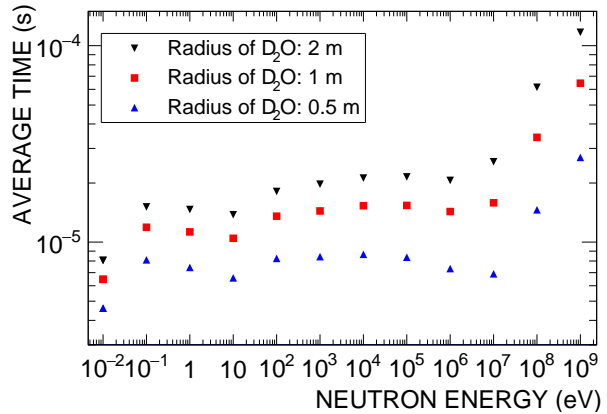}
  \caption{\label{fig:ave_times} Average time period a neutron spends inside the ion beam pipe for the different moderator sizes. The energy on the x-axis corresponds to the lower edge of the energy interval. Each interval corresponds to one energy decade, hence, the point at 10$^{8}$~eV corresponds to neutron energies between 10$^{8}$~eV and 10$^{9}$~eV.}
\end{center}
\end{figure}

FIG.~\ref{fig:ave_times} shows the average time period a neutron spends in the ion beam pipe as a function of its original energy. The period is only weakly dependent on the original energy since the neutrons are quickly moderated and trapped inside the moderator until they finally escape. If the energy exceeds the neutron separation energy of deuterium, about 2.2~MeV, additional neutrons are available. Their time periods inside the beam pipe are added to the time of the original neutron. The larger the moderator volume, the longer the time period inside the ion beam pipe. 

It is very important to note that the neutrons spend on average several microseconds in the ion beam pipe even for a relatively small moderator of only 50~cm in radius. A moderator of 2-m radius leads to a period of about 20~$\upmu$s. The longer the average time period inside the beam pipe, the higher the density of the neutron target for a given proton current.

\subsection{Protons energies of 800~MeV and 20~GeV}
\label{protons}

We simulated the setup with proton energies of 800~MeV and 20~GeV. The chosen energies correspond to the energies the LANSCE accelerator at Los Alamos National Laboratory (800~MeV)~\cite{LBR90} and the proton synchrotron at CERN (20~GeV)~\cite{GTB13,RLK14} deliver. We investigated two different tungsten target sizes, a cylinder with a radius of 1.5~cm and a length of 10 cm, and one with radius of 2.5~cm and a length of 50 cm. The four different simulation settings are listed in TABLE~\ref{Tab:neutron_yield}. The table also gives the neutron yield per proton. 

\begin{figure}
\begin{center}
\renewcommand{\baselinestretch}{1}
\small\normalsize
  \includegraphics[width=.49\textwidth]{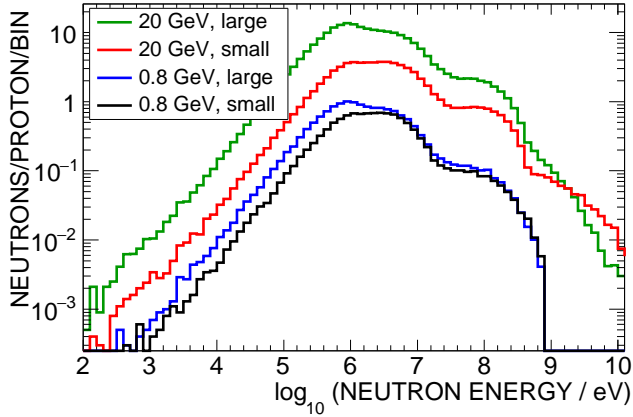}
  \caption{\label{fig:neutron_spectra_emitted} Spallation neutron spectra for the different settings, see also TABLE~\ref{Tab:neutron_yield}. The x-axis has a logarithmic binning with 10 bins per energy decade.}
\end{center}
\end{figure}

FIG.~\ref{fig:neutron_spectra_emitted} shows the energy spectra of the produced neutrons for all simulated combinations of target size and proton energy. In general, the number of spallation neutrons increases with the proton energy. As the 20-GeV protons will not be slowed down below the spallation threshold within 10~cm of tungsten, the neutron yield is significantly higher for the larger tungsten target. 

\begin{table}[htb]
 \caption{Neutron yield for the 4 different settings.}
   \label{Tab:neutron_yield}
   \renewcommand{\arraystretch}{1.5} 
   \begin{ruledtabular}
   \begin{tabular}{ccc}
    Proton Energy      & W-target                      & neutrons/proton \\
                       & radius $\times$ length               &           \\
       (GeV)           & (cm $\times$ cm)                     &           \\
    \hline
    0.8                & 1.5 $\times$ 10                      & $11.8$ \\
    0.8                & 2.5 $\times$ 50                      & $15.8$ \\
    20                 & 1.5 $\times$ 10                      & $70.1$ \\
    20                 & 2.5 $\times$ 50                      & $227$ \\  
   \end{tabular}
   \end{ruledtabular}
\end{table}

Each proton produces a certain amount of neutrons (see TABLE~\ref{Tab:neutron_yield}). These neutrons travel through the setup, are moderated by the heavy water, and eventually pass the ion beam pipe. The total time period $t_{\mathrm{neutron,tot}}$ sums up the time periods that all neutrons produced by a proton spend inside the beam pipe. Averaging over all protons, we obtain the average total time period that all neutrons produced by a proton spend inside the ion beam tube $\bar{t}_{\mathrm{neutron,tot}}$. To estimate the average number of neutrons inside the beam tube, $\bar{n}_{\mathrm{neutron}}$, we multiply $\bar{t}_{\mathrm{neutron,tot}}$ by the number of protons hitting the spallation target per time (proton current/elementary charge):

\begin{equation}\label{eq:neutrons_in_target}
  \bar{n}_{\mathrm{neutron}} = \frac{I_{\mathrm{proton}}}{e} \bar{t}_{\mathrm{neutron,tot}}
\end{equation}

\begin{figure}
\begin{center}
\renewcommand{\baselinestretch}{1}
\small\normalsize
  \includegraphics[width=.49\textwidth]{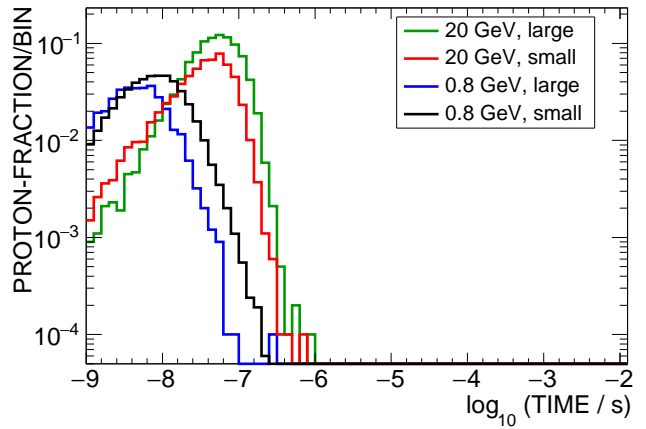}
  \caption{\label{fig:fast_time_spectra_geometry} Proton fraction as a function of the total time period the produced neutrons spend in the ion beam pipe. The simulations were carried out {\bf without a moderator}. The distributions are shown for different proton energies and tungsten target sizes (see also TABLE~\ref{Tab:neutron_times} and discussion of Equation~\ref{eq:neutrons_in_target}). The x-axis has a logarithmic binning with 10 bins per time decade.}
\end{center}
\end{figure}

\begin{figure}
\begin{center}
\renewcommand{\baselinestretch}{1}
\small\normalsize
  \includegraphics[width=.49\textwidth]{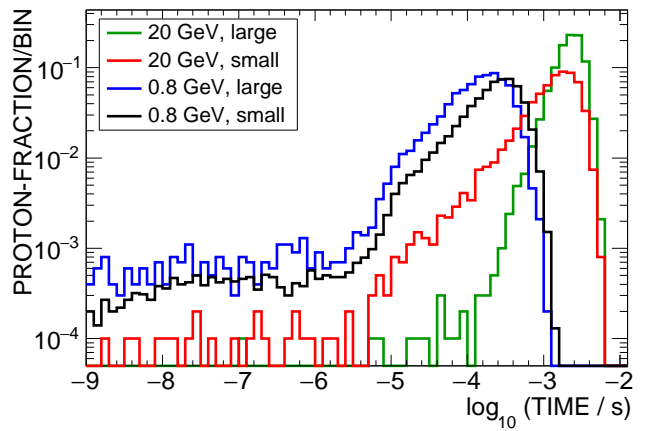}
  \caption{\label{fig:time_spectra_geometry} The distribution of the total time period that all neutrons produced by a single proton spend inside the ion beam pipe for a moderator 1~m in radius and all target/beam combinations listed in TABLE~\ref{Tab:neutron_times} and discussion of Equation~\ref{eq:neutrons_in_target}. The x-axis has a logarithmic binning with 10 bins per time decade. 
   }
\end{center}
\end{figure}

FIG.~\ref{fig:fast_time_spectra_geometry} shows ${t}_{\mathrm{neutron,tot}}$ for primary protons with different energies and tungsten target sizes, but without a moderator. The averages of the distributions corresponds to $\bar{t}_{\mathrm{neutron,tot}}$. Protons with an energy of 20~GeV produce more neutrons (see TABLE~\ref{Tab:neutron_yield}). Hence, the total time period the neutrons spend inside the beam pipe is higher compared to the 800-MeV protons. The 20-GeV protons produce about a factor of three more neutrons in the large tungsten target than in the small one. The 800-MeV protons produce only slightly more neutrons in the large target. But then the larger tungsten target acts as a neutron trap. The larger neutron production outweighs the neutron captures only in the case of 20-GeV protons, but not in the case of 800-MeV protons. Therefore, the total time period the produced neutrons spend in the ion beam pipe is shifted to smaller values for 800-MeV protons impinging on the large target. 

If a moderator is included, the total time period neutrons spend inside the ion beam pipe changes dramatically. FIG.~\ref{fig:time_spectra_geometry} shows the results for a moderator with a radius of 1~m. The averaged time period all neutrons produced per incident proton spend inside the ion beam pipe ($\bar{t}_{\mathrm{neutron,tot}}$) is listed in TABLE~\ref{Tab:neutron_times} for all combinations of proton energy, tungsten target size and moderator radius. 

\begin{table}[htb]
 \caption{Average total time perdiod neutrons spend inside the ion beam pipe per proton ($\bar{t}_{\mathrm{neutron,tot}}$, see Equation~\ref{eq:neutrons_in_target}). The recommended settings for a beam energy of 800~MeV and 20~GeV are printed in bold.}
   \label{Tab:neutron_times}
   \renewcommand{\arraystretch}{1.5} 
   \begin{ruledtabular}
   \begin{tabular}{cccc}
    Proton Energy      & W-target           & Moderator           & Time \\
                       & radius x length    & radius              &      \\
       (GeV)           & (cm x cm)          & (m)                 &  ($\upmu$s) \\
    \hline
    0.8                & 1.5 x 10           & 0.0                 & $0.0051$ \\
    0.8                & 1.5 x 10           & 0.5                 & $81$ \\
    0.8                & 1.5 x 10           & 1.0                 & $162$ \\
{\bf    0.8   }        & {\bf 1.5 x 10}     & {\bf 2.0 }          & {\bf 245 }\\ 
    0.8                & 2.5 x 50           & 0.0                 & $0.0022$ \\
    0.8                & 2.5 x 50           & 0.5                 & $62$ \\
    0.8                & 2.5 x 50           & 1.0                 & $142$ \\
    0.8                & 2.5 x 50           & 2.0                 & $224$ \\
    20                 & 1.5 x 10           & 0.0                 & $0.028$ \\
    20                 & 1.5 x 10           & 0.5                 & $490$ \\
    20                 & 1.5 x 10           & 1.0                 & $1000$ \\
    20                 & 1.5 x 10           & 2.0                 & $1560$ \\
    20                 & 2.5 x 50           & 0.0                 & $0.057$ \\
    20                 & 2.5 x 50           & 0.5                 & $1060$ \\
    20                 & 2.5 x 50           & 1.0                 & $2260$ \\
{\bf     20 }          & {\bf 2.5 x 50 }    & {\bf 2.0 }          & {\bf 3470} \\
\end{tabular}
\end{ruledtabular}
\end{table}

\begin{figure}
\begin{center}
\renewcommand{\baselinestretch}{1}
\small\normalsize
  \includegraphics[width=.49\textwidth]{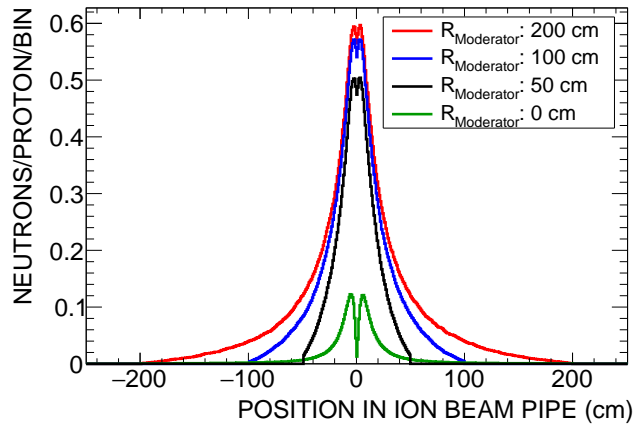}
  \caption{\label{fig:position_in_beam_pipe} Neutron position along the ion beam for different moderators, the small version of the tungsten target and 800-MeV protons. The x-axis has a bin width of 1~cm.}
\end{center}
\end{figure}

\begin{figure}
\begin{center}
\renewcommand{\baselinestretch}{1}
\small\normalsize
  \includegraphics[width=.49\textwidth]{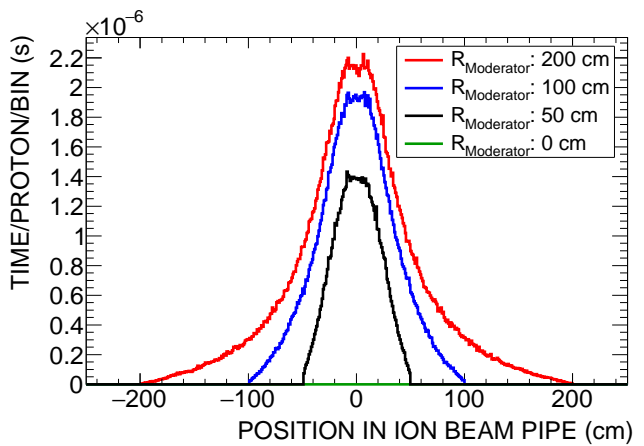}
  \caption{\label{fig:position_in_beam_pipe_time_weighted} The total time period neutrons spend inside the ion beam pipe as a function of position along the ion beam for different moderators, the small tungsten target and 800~MeV protons. Each entry corresponds to the sum of neutrons weighted with the time period the neutron has spent inside the tube. The x-axis has a bin width of 1~cm.}
\end{center}
\end{figure}

FIG.~\ref{fig:position_in_beam_pipe} shows the position distribution of neutrons inside the ion beam pipe. For this plot, the beam axis was sub-divided into discs of 1~cm width in the direction of the beam. Whenever a neutron entered one of the discs, it was recorded in the histogram.
The fact that high-energy neutrons don't spend much time inside the beam pipe is considered in FIG.~\ref{fig:position_in_beam_pipe_time_weighted}, where each entry was weighted with the corresponding time the neutron spend in the disc. From the comparison of Figs.~\ref{fig:position_in_beam_pipe} and \ref{fig:position_in_beam_pipe_time_weighted} one finds that the suppressed high-energy neutrons, which are usually coming directly from the spallation target, can be found mostly in the center of the ion beam pipe, close to the proton beam pipe. 

\begin{figure}
\begin{center}
\renewcommand{\baselinestretch}{1}
\small\normalsize
  \includegraphics[width=.49\textwidth]{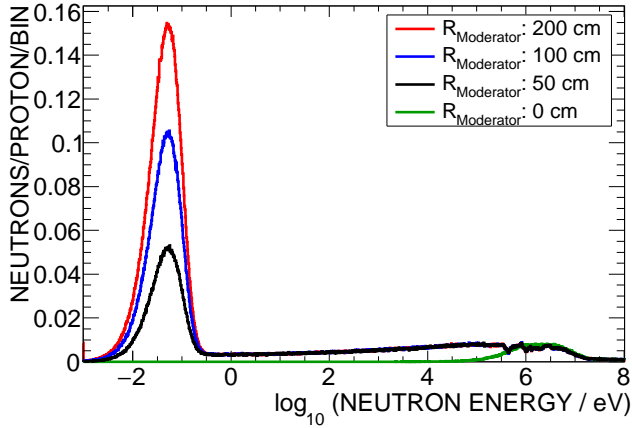}
  \caption{\label{fig:energy_in_beam_pipe} Energy distribution of all neutrons entering the ion beam pipe for different moderator sizes, the small tungsten target and 800-MeV protons. The x-axis has a logarithmic binning with 100 bins per energy decade.}
\end{center}
\end{figure}

\begin{figure}
\begin{center}
\renewcommand{\baselinestretch}{1}
\small\normalsize
  \includegraphics[width=.49\textwidth]{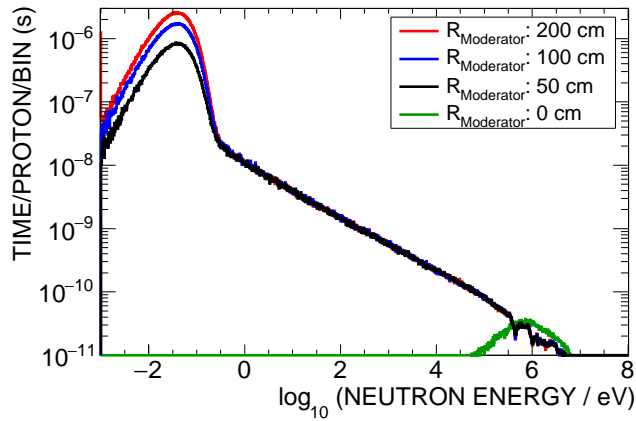}
  \caption{\label{fig:energy_in_beam_pipe_time_weighted} Total time period neutrons spend inside the ion beam pipe as a function of the neutron energy for different moderator sizes, the small tungsten target and 800-MeV protons. Each entry corresponds to the sum of neutrons weighted with the time period the neutron spent inside the ion tube. The x-axis has a logarithmic binning with 100 bins per energy decade.}
\end{center}
\end{figure}

FIG.~\ref{fig:energy_in_beam_pipe} shows the energy distribution of the neutrons in the ion beam pipe the small tungsten target, 800-MeV protons and different moderator sizes. Without a moderator, the energy distribution resembles the spallation neutron spectrum, compare FIG.~\ref{fig:neutron_spectra_emitted}. The moderation process produces neutron spectra with a thermal energy distribution. The larger the moderator, the more neutrons per proton enter the ion beam pipe. In FIG.~\ref{fig:energy_in_beam_pipe_time_weighted} shows the energy spectra are weighted with the corresponding time period the neutron spent inside the ion beam pipe. High-energy neutrons pass the pipe quickly. The moderated neutrons spend up to 2~$\upmu$s in the pipe.  

As can be seen from FIGs.~\ref{fig:energy_in_beam_pipe} and~\ref{fig:energy_in_beam_pipe_time_weighted}, the neutron spectrum in the ion beam pipe is clearly dominated by low-energy neutrons. Therefore, the center-of-mass energy of the ion-neutron-collision in realistic experiments will be defined by the energy of the ions. The ion energy can easily be tuned in the storage ring by applying, for instance, electron cooling at a defined beam energy.

\begin{figure}
\begin{center}
\renewcommand{\baselinestretch}{1}
\small\normalsize
  \includegraphics[width=.49\textwidth]{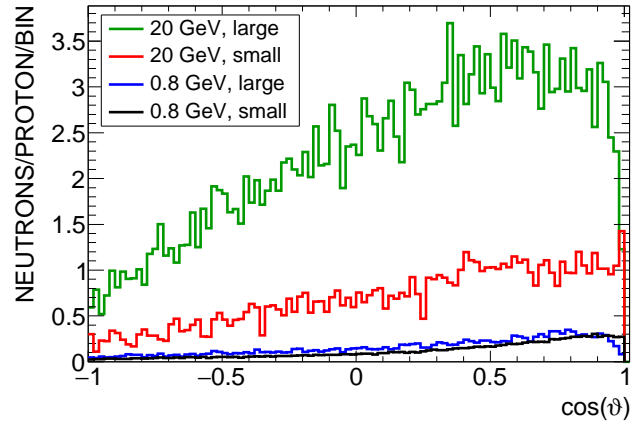}
  \caption{\label{fig:cos_theta} Neutron emission angle from the spallation target with respect to the direction of the protons for  all settings. The x-axis has 100 bins.}
\end{center}
\end{figure}

The simulations were repeated with a different tungsten target position relative to the moderator. We moved the target upstream since most of the neutrons are emitted in the direction of the impinging high-energy proton beam (FIG.~\ref{fig:cos_theta}), potentially gaining more forward-emitted neutrons. However, the average time period neutrons spend inside the ion beam pipe was reduced. The scattering processes in the moderator alters the original direction of the neutron quickly. Hence, the emission angle of the neutrons is not important.

\section{Possible realizations}\label{realizations}

We discuss possible realizations of the proposed setup at different experimental facilities. We consider the available proton currents and energies to estimate the number of neutrons inside an ion beam tube running through a heavy-water moderator. We estimate the average time period neutrons spend inside the ion beam tube per incoming proton from the proton energy, TABLE~\ref{Tab:neutron_times} and Equation~\ref{eq:neutrons_in_target}. We calculate the areal neutron density $\eta_{\mathrm{neutron}}$ by dividing the average number of neutrons inside the beam tube $\bar{n}_{\mathrm{neutron}}$ by the cross section of the ion beam pipe $A_{\mathrm{ion\;pipe}}$:

\begin{equation}\label{eq:neutrons_in_target_areal}
  \eta_{\mathrm{neutron}} = \frac{\bar{n}_{\mathrm{neutron}}}{A_{\mathrm{ion\;pipe}}} = \frac{I_{\mathrm{proton}}}{e} \frac{\bar{t}_{\mathrm{neutron,tot}}}{A_{\mathrm{ion\;pipe}}}
\end{equation}

The set of simulations described here was driven by the specific opportunities of the LANSCE-accelerator at LANL and the PS at CERN. These facilities have already well-established spallation neutron sources with the corresponding driving accelerators. Therefore, they are currently the most promising options. However, the idea of using a heavily moderated spallation source in conjunction with a storage ring is certainly not restricted to these facilities, but may be setup at other institutions like GSI/FAIR, HIRFL, HIAF, NSCL/MSU, MYRRHA. In particular, new developments in cyclotron technology like fixed field alternating gradient (FFAG) machines \cite{CaM15} offer the possibility to realize this setup at a dedicated facility.  

\subsection{Los Alamos National Laboratory (LANL)}

The Lujan target at LANSCE/LANL operates with an average proton current of about 100~$\upmu$A. We consider an ion beam tube with a cross section of 20~cm$^{2}$. We obtain a neutron density of about $8 \times 10^{9}$~n/cm$^2$ using the largest moderator with a diameter of 200~cm. The neutron density is only a factor of two less than described in reference~\cite{ReL14}, where a neutron flux of $10^{14}$~n/cm$^2$/s in a reactor and an interaction length of 0.5~m was assumed. The results for the different moderator sizes are given in TABLE~\ref{Tab:neutron_density}.

\begin{table}[htb]
 \caption{Neutron density for the simulated setup at two facilities: 100~$\upmu$A proton beam with 800~MeV and the small tungsten target (LANL), as well as $3\times 10^{12}$ protons/s with an energy of 20~GeV and the large tungsten target (CERN), see Equation~\ref{eq:neutrons_in_target_areal}. 
         }
   \label{Tab:neutron_density}
   \renewcommand{\arraystretch}{1.5} 
   \begin{ruledtabular}
   \begin{tabular}{ccc}
   Moderator radius          & \multicolumn{2}{c}{Neutron density (cm$^{-2}$)} \\
   (m)                 &  LANL               &  CERN             \\
   \hline                            
   0.0                 & $1.6\times 10^6$    & $8.7\times 10^3$  \\
   0.5                 & $2.6\times 10^9$    & $1.6\times 10^8$  \\
   1.0                 & $5.2\times 10^9$    & $3.6\times 10^8$  \\
   2.0                 & $7.8\times 10^9$    & $5.4\times 10^8$  \\
		  		  
   \end{tabular}
   \end{ruledtabular}
\end{table}

The LANSCE accelerator provides H$^-$ and H$^+$ ions~\cite{LBR90,MiW90}. A magnet bends H$^-$ ions to the southern experimental areas, and H$^+$ ions to the northern area. Additional facilities, like the proposed combination of an ion storage ring and a neutron target, could be installed at the northern area. Here, H$^+$ beams with currents up to 1~mA could be delivered. In particular, the nearby isotope production facility (IPF), located at the beginning of the LANSCE accelerator, makes this idea very attractive. The required amount of radioactive material for an experiment in inverse kinematics is orders of magnitude {\bf smaller} than the amount of material needed for a traditional time-of-flight experiment like DANCE~\cite{ERB08, WBC15, CoR07}. Typically, the storage ring needs to be filled every few minutes with about 10$^{7}$ ions \cite{MAB15} consuming about about $1.5\times 10^{10}$ atoms during a day.

\subsection{European Organization for Nuclear Research (CERN)}

The n\_TOF/ISOLDE experiments receive about \mbox{$3\times 10^{12}$} protons/s with an energy of 20~GeV. On average, these protons could produce between $1.5\times 10^{9}$ and $10^{10}$ neutrons in an ion beam pipe as described here. The resulting neutron densities of up to $6\times 10^8$~n/cm$^2$ in the pipe are listed in TABLE~\ref{Tab:neutron_density}. The installation of the well-suited ion storage ring TSR at HIE-ISOLDE \cite{GLR12} would provide the main ingredients of the proposed setup.

\subsection{GSI Helmholtz Center for Heavy Ion Research (GSI) and Facility for Antiproton and Ion Research (FAIR)}

The universal linear accelerator UNILAC and the \mbox{18-T$\cdot$m} heavy-ion synchrotron SIS-18 are the driver accelerators at GSI in Darmstadt (Germany). Proton intensities of up to $2.1\times 10^{11}$ protons per pulse have routinely been achieved \cite{BAA16}. The magnetic rigidity of 18~T$\cdot$m allows to accelerate protons to energies of 4.5~GeV. The 100-T$\cdot$m synchrotron of the future FAIR facility~\cite{FAIR} will provide about $5\times 10^{12}$ protons/s with an energy of 28.8~GeV.

A variety of specialized facilities like fragment separators \cite{GAB92,GWW03}, storage rings \cite{Fra87, LAA16},  experimental caves, and beam lines \cite{BLS13, SLB14} provides numerous opportunities to realize a low-energy storage ring combined with a spallation target described here.

\subsection{Heavy Ion Research Facility in Lanzhou (HIRFL) and Cooler Storage Ring (CSR) Complex (IMP) and High Intensity Heavy-Ion Accelerator Facility (HIAF)}

The Heavy Ion Research Facility in Lanzhou (HIRFL), China, which is similar to GSI, has been operational since 2007~\cite{XZW02}. The maximum magnetic rigidity of the main synchrotron ring is 10.6~T$\cdot$m (maximum proton energy is 2.3~GeV). The existing cooler storage ring CSRe might be employed to store radioactive ions. The High Intensity Heavy-Ion Accelerator Facility (HIAF) has been approved and will be constructed in Huizhou, China. The considered rigidity of the main synchrotron will be around 30~T$\cdot$m, thus providing about 8~GeV protons. The beam intensities are comparable to those expected at FAIR. Several high-energy as well as low-energy storage rings are being considered for HIAF. Here, the proposed combination of a storage ring with a spallation target could be considered already at the planning stage of the facility. Furthermore, a high power 1.0 to 1.5~GeV dedicated proton linear accelerator is under construction at the same location as HIAF within the running Accelerator-Driven Systems (ADS) project in China.

\section{Conclusions}

The combination of an intense neutron source and an ion storage ring would be unique for a direct measurement of neutron-induced reactions, as already discussed in~\cite{ReL14}. Direct kinematics, where the neutrons impinge on the target of interest, require the measurement of the light reaction products. The $\gamma$-rays from the investigated reaction have to be discriminated from the sample decay $\gamma$-rays. The proposed technique is based on the detection of the heavy, projectile-like residues of the reactions, which is a significant advantage for (a) the measurement of (n,$\gamma$) cross sections of $\gamma$-emitting samples or of fissile nuclei, (b) the measurement of (n,p) and (n,$\alpha$) cross sections at low energies, as there is no need to detect the emitted protons and $\alpha$-particles, and (c) for cross section measurements on relatively long-lived or even stable nuclei.

\section{Summary}

A typical spallation neutron source can be modified to build a neutron target by combination with a large moderator of heavy water. The resulting neutron target, intercepted by an ion beam, can be used to investigate neutron-induced reactions. This technique is of advantage if the sample size in a traditional time-of-flight setup is limited, either because of the decay properties of the investigated isotope or because of the range of the reaction products. In particular, the combination with a storage ring makes this techniques feasible. Existing facilities like LANSCE at LANL, n\_TOF/ISOLDE at CERN, or GSI/FAIR could be complemented with the setup described here. 

\begin{acknowledgments}

This research has received funding from the European Research Council under the European Unions's Seventh Framework Programme (FP/2007-2013) / ERC Grant Agreement n. 615126. We thank Hushan Xu for his valuable information about the planned spallation sources in China (ADS and HIAF).

\end{acknowledgments}

\bibliography{/home/reifarth/Texte/paper/refbib}
\bibliographystyle{apsrev4-1}

\end{document}